\documentclass[english,a4paper,twocolumn,amsmath,amssymb,superscriptaddress,tightenlines,reprint]{revtex4-2}
\usepackage{bm}
\usepackage{graphicx,xcolor}
\definecolor{linkcolor}{HTML}{004191}
\usepackage[colorlinks,linkcolor=linkcolor,citecolor=linkcolor, urlcolor=linkcolor]{hyperref}
\usepackage{physics}
\usepackage{mlmodern}
\usepackage{amsmath, amssymb}
\usepackage{mathtools}

\newcommand{\Int}[1]{\int\dd #1\;}
\newcommand{\IInt}[3]{\int_{#2}^{#3}\dd #1\;}

\newcommand{\al}{\alpha}

\newcommand{\eps}{\varepsilon}
\newcommand{\kap}{\kappa}

\newcommand{\sig}{\sigma}

\newcommand{\Om}{\Omega}

\newcommand{\x}{\bm r}
\newcommand{\kT}{k_\text{B}T}

\begin{document}

\title{Impact of currents on non-equilibrium coexistence in chemically driven mixtures}

\author{Ellen Meyberg}
\affiliation{Institute for Theoretical Physics IV, University of Stuttgart, Heisenbergstr. 3, 70569 Stuttgart, Germany}
\author{Joshua F. Robinson}
\affiliation{STFC Hartree Centre, Sci-Tech Daresbury, Warrington, WA4 4AD, United Kingdom}
\affiliation{H.\ H.\ Wills Physics Laboratory, University of Bristol, Bristol BS8 1TL, United Kingdom}
\author{Thomas Speck}
\affiliation{Institute for Theoretical Physics IV, University of Stuttgart, Heisenbergstr. 3, 70569 Stuttgart, Germany}

\begin{abstract}
    Virtually every biological function emerges through the organization of molecules in time and space. Consequently, a major challenge in statistical physics is to uncover the universal principles governing macromolecular self-organization within the crowded, non-equilibrium environment of the cell. Here, we investigate a class of models where molecules maintain a conserved total concentration but can switch ``identities'', thereby modulating their intermolecular interactions. By enforcing thermodynamic consistency via the local detailed balance condition, we derive the steady-state criteria determining coexisting concentrations in a binary mixture. For non-constant transition rates and using a sharp-interface approximation, we obtain jump conditions that generalize Gibbs' coexistence criteria of equal pressure and chemical potential. We demonstrate that these jumps balance the chemical potential differences of individual species against their currents, which are confined to the interfacial region.
\end{abstract}

\maketitle


\section{Introduction}

In his seminal work, Gibbs laid the foundation for thermodynamics to move beyond simple substances and to describe multicomponent mixtures~\cite{gibbs78}. To deal with systems where the composition can change, he established that for a mixture to exist in stable coexistence, the system must satisfy not only thermal and mechanical uniformity but also the equality of chemical potential for every constituent across distinct phases. No comparable general framework exists for systems steadily driven away from equilibrium, where fluxes of energy and matter prevent a simple minimization of free energy. In these contexts, phase behavior is governed by kinetic competition and dissipative processes. This shift from static coexistence to dynamic organization is particularly evident in biomolecular condensates, such as nucleoli or stress granules, which form via liquid-liquid phase separation and coexist with the surrounding cytoplasm or nucleoplasm~\cite{hyman14,banani17,boeynaems18,alberti25}. The formation, maintenance, and dissolution of these cellular compartments are frequently linked to energy-dissipating chemical reaction cycles~\cite{oflynn21,zwicker22,tayar23}.

Theoretical models for the dynamic evolution of concentrations naturally assume the form of reaction-diffusion systems~\cite{weber19,alston22,hafner24,kirschbaum21,julicher24,bauermann25a}, which have a long and rich history starting with Turing's seminal work on morphogenesis~\cite{turing52,kondo10,krause21}. For idealized models of mixtures neglecting intermolecular interactions while employing non-linear reactive fluxes, Frey and coworkers have emphasized the crucial role of \emph{conservation laws} on pattern formation~\cite{halatek18a,weyer25,frey26}. The presence of a conservation law in combination with free diffusion confines interfaces to a ``flux-balance'' plane in chemical space within which diffusive and reactive fluxes balance in steady state, providing a geometric framework to construct interfaces and patterns~\cite{brauns20}. Going beyond ideal mixtures and taking into account intermolecular interactions~\cite{li20,avanzini21,miangolarra23}, recent results based on stochastic thermodynamics~\cite{seifert12,rao16,seifert25} and large deviation theory emphasize that in the presence of interactions not only the transition \emph{rates} but the reactive \emph{fluxes} need to obey the local detailed balance condition to be thermodynamically consistent~\cite{falasco25}. While local detailed balance has been incorporated into molecular dynamics~\cite{berthin25,zippo25,lavagna25} and lattice gas simulations~\cite{cho25}, the systematic exploration of its theoretical ramifications for phase segregation remains in its early stages~\cite{avanzini21,avanzini24a}.

Although observables such as coexisting concentrations can be extracted numerically from specific models, a deeper theoretical understanding on the existence and role of the chemical potential in chemically driven mixtures is lacking. In particular the conditions when particle currents are possible, and how these currents impact the coexistence criteria, have remained unclear. Guided by the two principles of mass conservation and local detailed balance, here we develop the underlying generalized coexistence criteria for \emph{non-ideal} chemically driven binary mixtures. Such non-ideal reaction-diffusion systems are still confined to a flux-balance space, but the geometry of this space is no longer a simple plane. We argue that spatial currents are confined to the interface and imply a jump of chemical potential across the interface. Specifically, we study a schematic toy model that exposes general features found in cellular processes: Molecules can change and undergo transitions between ``species'', e.g., molecular conformations or post-translational modifications such as phosphorylation and methylation~\cite{li22}. The molecules diffuse and interact, and these interactions depend on their current species. And finally, the total concentration is conserved, i.e., we focus on time scales on which the creation and degradation of molecules can be ignored.


\section{Results and discussion}

\subsection{The system}

\begin{figure}[t]
    \centering
    \includegraphics{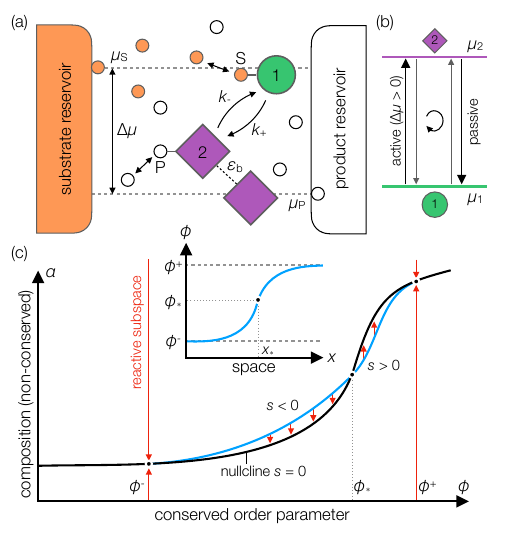}
    \caption{Chemically active binary mixture. (a)~Sketch of macromolecules that can transition between two ``species'' (disk and oblique square). There are two reaction channels: spontaneous passive transitions and activated transitions involving the conversion of a molecular solute from substrate (S) to product (P) with rates $k^\nu_\pm$ ($\nu=\text{a,p}$). The concentrations of the molecular solutes are assumed to be fixed through reservoirs at constant chemical potentials. (b)~Steady-state chemical fluxes at constant concentrations: Active transitions are biased towards species $2$ (for $\Delta\mu>0$), the chemical potential of which is increased. Consequently, passive transitions can decrease the free energy converting species 2 back to 1 so that overall the concentrations remain constant. (c)~Chemical space spanned by the total concentration $\phi$ of macromolecules and their mole fraction $\al$. Along the nullcline $\bar\al(\phi)$ (solid black line) the net flux vanishes. In a uniform system the dynamics is confined to the reactive subspaces at constant $\phi$ (red lines). The blue line sketches the non-linear flux-balance space for an inhomogeneous system connecting both bulk phases with $\phi^\pm$ on the nullcline. The inset shows the corresponding profile $\phi(x)$ in real space. While the flux-balance space coincides with the nullcline in passive systems obeying detailed balance, driving the mixture away from equilibrium might displace the flux-balance space implying currents across the interface (thin red arrows).}
    \label{fig:system}
\end{figure}

We are interested in suspensions of macromolecules with their total number conserved. Macromolecules can either be found in different conformations or their chemical structure can be modified, e.g., through enzymes. Macromolecules are thus divided into chemical ``species'' (M$_k$) with stochastic transitions between them. Importantly, the intermolecular interactions between macromolecules depend on their species. In the following, we will focus on the simplest case of just two species. The transition
\begin{equation}
    \text{S} + \text{M}_1 \xrightleftharpoons[k^\text{a}_-]{k^\text{a}_+} \text{M}_2 + \text{P}
    \label{eq:active}
\end{equation}
between these species is ``activated'' in the sense that it involves the conversion of molecular solutes [in the following designated as substrate (S) and product (P)] with transition rates $k^\text{a}_\pm$ [Fig.~\ref{fig:system}(a)]. We assume that the chemical potentials $\mu_\text{S,P}$ of molecular solutes are held fixed through chemiostats with difference $\Delta\mu\equiv\mu_\text{S}-\mu_\text{P}$. As a specific example, consider a substrate molecule binding to the macromolecule and its conversion (the barrier is lowered due to the binding) to product causes a conformational change of the macromolecule; but the scheme is also applicable to enzymatic reactions~\cite{cotton22,lavagna25}. Passive transitions between species are also possible in the absence of molecular solutes with rates $k^\text{p}_\pm$ obeying detailed balance. Biasing the active transitions through $\Delta\mu\neq 0$ leads to chemical fluxes [Fig.~\ref{fig:system}(b)] driving the system into a non-equilibrium steady state.

The concentration fields $\rho_k(\x,t)$ of the two conformations ($k=1,2$) evolve according to the continuity equations
\begin{equation}
    \partial_t\rho_k + \nabla\cdot\bm j_k = s_k
    \label{eq:rho}
\end{equation}
with particle currents $\bm j_k$. The reactive fluxes $r^\nu_\pm$ describe the number of macromolecules (per volume per time) that undergo the transition $1\to 2$ ($+$) and the reverse transition $2\to 1$ ($-$). The net fluxes into (out of) each conformation then obey $s_1=-s$ and $s_2=s$ with $s\equiv\sum_\nu(r^\nu_+-r^\nu_-)$ summing over all reaction channels $\nu=(\text{a,p})$. From Eq.~\eqref{eq:rho}, we see that the total concentration $\phi\equiv\rho_1+\rho_2$ obeys $\partial_t\phi+\nabla\cdot\bm j=0$ with particle current $\bm j\equiv\bm j_1+\bm j_2$, and thus $\phi$ is indeed a conserved quantity.

The underlying microscopic rates for spatial translations obey \emph{detailed balance}, which implies that particle currents $\bm j_k=-\rho_k\nabla\mu_k$ can only be driven through non-uniform chemical potentials $\mu_k=\fdv{F}{\rho_k}$. To keep the mathematical expressions tractable while exposing the relevant mechanisms, we make a few simplifications. First, we assume equal mobility for both species, which we absorb into the definition of time. Second, we assume the expression
\begin{equation}
    F = \Int{^3\x} \left[\frac{1}{2}\sum_{ij}\kap_{kl}(\nabla\rho_k)(\nabla\rho_l) + f(\bm\rho) \right]
\end{equation}
for the free-energy functional with point-wise free-energy density $f(\bm\rho)$ and including the lowest-order expansion in the concentration gradients. For the symmetric coefficients $\kap_{ij}$ we employ values so that the square-gradient term reduces to $(\kap/2)|\nabla\phi|^2$, i.e., only the change of the total concentration is penalized. The limit of ideal mixtures corresponds to dropping the square-gradient term ($\kap=0$) and employing the ideal gas free-energy density $f^\text{id}=\kT\sum_k\rho_k\ln(v_T\rho_k)$ with temperature $T$, which yields $\bm j^\text{id}_k=-\kT\nabla\rho_k$ ($v_T$ is some molecular volume and $k_\text{B}$ is Boltzmann's constant).

Instead of the concentrations $\rho_k$, it is often more convenient to employ $\phi$ and the mole fraction $0\leqslant\al\leqslant 1$ as variables so that $\rho_1=(1-\al)\phi$ and $\rho_2=\al\phi$. Exploiting the chain rule, for the chemical potentials we obtain
\begin{equation}
    \mu_1 = \mu - \al\epsilon, \qquad
    \mu_2 = \mu + (1-\al)\epsilon
    \label{eq:mu:k}
\end{equation}
with
\begin{equation}
    \mu \equiv \fdv{F}{\phi} = \pdv{f}{\phi} - \kap\nabla^2\phi, \qquad
    \epsilon \equiv \frac{1}{\phi}\pdv{f}{\al}
    \label{eq:mu}
\end{equation}
so that $\epsilon=\mu_2-\mu_1$ is the (local) difference between the chemical potentials of the two species.

While $\phi$ is a conserved quantity, the mole fraction $\al$ is not conserved. The kinetics of a uniform system is thus confined to a \emph{reactive phase space}, i.e., to lines along which $\phi$ is constant, and relaxes towards stable reactive equilibria characterized by vanishing net fluxes ($s=0$). The manifold of these equilibria is called the \emph{nullcline}, which in our case is simply a line $\al=\bar\al(\phi)$ parametrized by $\phi$ [Fig.~\ref{fig:system}(c)]. Nullcline and reactive phase spaces are geometric structures that organize chemical space. In inhomogeneous systems, different regions in space relax towards different reactive equilibria connected through interfaces.

\subsection{Local detailed balance and dissipated heat}

In the presence of \emph{intermolecular} interactions encoded in the free energy density $f(\al,\phi)$, the reactive fluxes (not just the bare transition rates $k_\pm$) need to obey the \emph{local detailed balance condition} (appearing in Refs.~\citenum{weber19,kirschbaum21} and derived in Ref.~\citenum{falasco25} using large deviation methods)
\begin{equation}
    \kT\ln\frac{r^\text{a}_+}{r^\text{a}_-} = -\epsilon + \Delta\mu
    \label{eq:ldb}
\end{equation}
to achieve thermodynamic consistency between chemical driving and the heat dissipated into the solvent kept at temperature $T$. Each transition of a macromolecule involves the transfer of a molecular solute between the two reservoirs, which enters Eq.~\eqref{eq:ldb} through the difference $\Delta\mu$ between the chemiostated species. Such a conversion does not change the local concentration $\phi$ and thus only the partial derivative of $f$ with respect to $\al$ enters Eq.~\eqref{eq:ldb}, cf. Eq.~\eqref{eq:mu}. In the following, we employ the fluxes
\begin{equation}
    r^{\nu}_\pm = \phi\sqrt{\al(1-\al)}k^{\nu}_\pm\exp\left\{\mp\frac{\epsilon}{2\kT}\right\}
    \label{eq:fluxes}
\end{equation}
through reaction channel $\nu$ with transition rates $k_\pm^\nu$ respecting $k_+^\text{a} / k_-^\text{a} = \exp(\Delta \mu / \kT)$ and $k_+^\text{p} = k_-^\text{p}$ to satisfy Eq.\ \eqref{eq:ldb}. These fluxes reduce to the ideal case for vanishing intermolecular interactions, i.e., $f\to f^\text{id}$ leading to the mass-action kinetics $r^\text{id}_+=k_+\rho_1$ and $r^\text{id}_-=k_-\rho_2$. We stress that for interacting mixtures fluxes necessarily divert from these mass-action kinetics. As an immediate consequence of local detailed balance, we can determine the value
\begin{equation}
    \bar\epsilon \equiv \kT\ln\frac{k^\text{a}_++k^\text{p}_+}{k^\text{a}_-+k^\text{p}_-}
    \label{eq:nc:eps}
\end{equation}
of $\epsilon$ on the nullcline through plugging Eq.~\eqref{eq:fluxes} into the expression for the net flux, setting $s=0$, and rearranging for $\epsilon$. Importantly, $\bar\epsilon$ depends only on the transition rates and thus relates the thermodynamic difference of chemical potentials to the kinetics of molecular transformations.

Following standard arguments from stochastic thermodynamics~\cite{seifert12}, the dissipation rate in steady state can be split into
\begin{multline}
    \dot Q_\text{c} = \kT\sum_\nu\Int{^3\x}(r^\nu_+-r^\nu_-)\ln\frac{r^\nu_+}{r^\nu_-} \\ = \Int{^3\x}\left[(r^\text{a}_+-r^\text{a}_-)(-\epsilon+\Delta\mu)-(r^\text{p}_+-r^\text{p}_-)\epsilon\right]
\end{multline}
due to the reactions and
\begin{multline}
    \dot Q_\text{d} = \sum_k\Int{^3\x}\bm j_k\cdot(-\nabla\mu_k) = \sum_k\Int{^3\x}(\nabla\cdot\bm j_k)\mu_k \\ = \Int{^3\x}s(\mu_2-\mu_1) = \Int{^3\x}s\epsilon
\end{multline}
due to non-vanishing particle currents. Here we have performed an integration by parts and used $\nabla\cdot\bm j_k=s_k$. The sum becomes
\begin{equation}
    \dot Q = \dot Q_\text{c} + \dot Q_\text{d} = \Delta\mu\Int{^3\x}(r^\text{a}_+-r^\text{a}_-) \equiv \mathcal J\Delta\mu.
    \label{eq:diss}
\end{equation}
As required by thermodynamic consistency, the total dissipation rate $\dot Q$ is indeed equal to the amount of free energy transferred between the reservoirs, i.e., the difference of chemical potential $\Delta\mu$ times the net number $\mathcal J$ of substrate molecules converted per time. Due to the tight coupling implied in Eq.~\eqref{eq:active}, this chemical flux between reservoirs is equal to the net flux $s^\text{a}\equiv r^\text{a}_+-r^\text{a}_-$ through the active channel.

\subsection{Steady-state coexistence}

Clearly, for $\Delta\mu=0$ the dissipation vanishes and the steady state corresponds to thermal equilibrium. Thermodynamics provides us with a set of conditions that unambiguously determine the coexisting phases: the chemical potentials $\mu_1=\mu_2$ have to be equal with $\mu=\mu_\text{coex}$ taking a constant value everywhere. At any point in space the system resides on the nullcline, $\al=\bar\al(\phi)$, with a spatial profile $\phi(\x)$ that connects the bulk concentrations $\phi^\pm$. In addition, the bulk pressures have to be equal (for a flat interface).

\begin{figure}[t]
    \centering
    \includegraphics{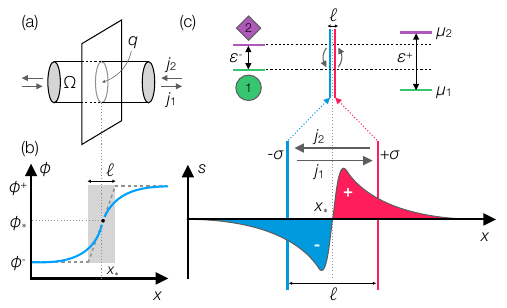}
    \caption{Active interface. (a)~Sample integration region $\Omega$ for the divergence theorem. A non-zero ``charge'' $q$ at the interface enclosed by $\Omega$ would imply bulk currents $\bm j_k$. (b)~Simplified interface constructed from the tangent at $\phi_\ast$, the intersection of which with the bulk concentrations $\phi^\pm$ determines the effective width $\ell$. (c)~Sketch of the net flux $s(x)$ across a charge-free interface. We replace the smooth function $s(x)$ by two planar sheets separated by a small distance $\ell$ with surface charges $\pm\sig$ adding to $q=0$. The charges drive lateral currents $\bm j_1=-\bm j_2=\sig\bm e_x$ between the plates. Zooming out, the interface becomes a dipole sheet and the potential difference $\epsilon$ has to jump across the interface.}
    \label{fig:interface}
\end{figure}

Before embarking on developing the coexistence criteria for $\Delta\mu\neq0$, we can already anticipate some qualitative features of the steady-state behavior from the formal analogy with electrostatics. In steady state, $\nabla\cdot\bm j_k=s_k$ [cf. Eq.~\eqref{eq:rho}] so that the currents take the role of the electric displacement field while $s$ is akin to a charge density. Since for two species $s_1=-s_2$ we conclude $\bm j_1=-\bm j_2$ so that the total steady-state current vanishes (technically there could be a solenoidal current, but there is no source for such a current in our system). For a subregion $\Om$ of the system, it is useful to introduce the ``charge'' $q_\Om=\IInt{^3\x}{\Om}{}s(\x)$. Let us consider an infinite plane corresponding to a flat interface between two coexisting phases. We choose $\Om$ as, e.g., a cylinder aligned with the interface normal so that its bases are within both bulk phases [Fig.~\ref{fig:interface}(a)]. A non-zero enclosed charge $q_\Om$ then implies non-vanishing \emph{bulk} currents $\bm j_k\neq 0$. Since within reactive subspaces the system relaxes towards the nullcline, we conclude that in bulk the system still resides on the nullcline and thus $s=0$. Within the interface, where concentrations are changing, potentially $s\neq 0$ but its integral through the interface connecting the bulk phases has to vanish.

Employing Eq.~\eqref{eq:mu:k}, the expression
\begin{equation}
    \bm j = -\sum_k\rho_k\nabla\mu_k = -\phi\nabla\mu + \phi\epsilon\nabla\al
    \label{eq:j}
\end{equation}
for the total current depends on the gradients of the chemical potential $\mu$ and the mole fraction $\al$. Moreover, exploiting $\bm j=0$ the currents of the single species can be written
\begin{equation}
    \bm j_2 = -\al(1-\al)\phi\nabla\epsilon = -m\nabla\epsilon = -\bm j_1,
\end{equation}
where $m\equiv\al(1-\al)\phi$ is an effective mobility. Using once more the vocabulary of electrostatics, $m$ corresponds to the permittivity and $\epsilon$ to the electric potential. Clearly, currents are driven by the spatial gradient of $\epsilon$, and taking the divergence yields the Poisson equation
\begin{equation}
    -\nabla\cdot(m\nabla\epsilon) = s
    \label{eq:eps}
\end{equation}
so that any gradient of $\epsilon$ requires non-zero reactive fluxes.

As a first central result, we conclude that for uniform $\epsilon$ all particle currents vanish and Eq.~\eqref{eq:j} tells us that the coexisting phases follow from the conventional Gibbs criteria but involving the \emph{rebalancing potential} $\eta(\phi)\equiv\mu-\bar\al\epsilon$, which in steady state acquires a constant value $\eta_\text{coex}$. We stress that for $\Delta\mu\neq 0$ the mixture is out of equilibrium with a non-zero dissipation rate $\dot Q\neq 0$ [Eq.~\eqref{eq:diss}], but the reactive fluxes balance locally so that $s=0$ without involving lateral particle currents~\cite{avanzini24a}. Since the system is confined to the nullcline everywhere, from Eq.~\eqref{eq:nc:eps} we further conclude that this situation corresponds to concentration-independent transition rates $k^\nu_\pm$.

\subsection{Pressure}

In order to determine the value of $\eta_\text{coex}$ at coexistence, we require a second condition. To this end, we note that the total current Eq.~\eqref{eq:j} can be rearranged differently to
\begin{equation}
    \bm j = -\nabla(\phi\mu-f) - \kap(\nabla^2\phi)\nabla\phi
\end{equation}
using $\nabla f=(\partial_\phi f)\nabla\phi+\phi\epsilon\nabla\al$ together with Eq.~\eqref{eq:mu}, where we recognize $p=\phi\mu-f$ as the thermodynamic (bulk) pressure. Spelling out the interfacial term in spherical coordinates leads to
\begin{multline}
    (\nabla^2\phi)\nabla\phi = \left(\partial_r^2\phi+\frac{2}{r}\partial_r\phi\right)(\partial_r\phi)\bm e_r \\ = \left(\partial_r\frac{1}{2}|\partial_r\phi|^2 + \frac{2}{r}|\partial_r\phi|^2\right)\bm e_r.
\end{multline}
Setting again $\bm j=0$, we find the required second condition
\begin{equation}
    \partial_rP = -\frac{2\kap}{r}|\partial_r\phi|^2, \qquad P = \frac{1}{2}\kap|\partial_r\phi|^2 + \phi\mu - f.
    \label{eq:eff_pressure}
\end{equation}
For a flat interface $\partial_rP=0$ and we can equate the bulk pressures, $p(\phi^+)-p(\phi^-)=p_\text{coex}$, while for a spherical droplet the right-hand term captures the usual interfacial tension. While we have thus recovered the expected thermodynamic pressure, importantly, this derivation extends to non-vanishing currents $\bm j_k\neq 0$.

\subsection{Generalized coexistence criteria}

Non-vanishing currents require that the transition rates $k^\nu_\pm$ depend on $\phi$ since then the difference $\epsilon$ between chemical potentials is no longer uniform. To proceed and to obtain further analytical insight without knowledge of the actual profile $\phi(\x)$, we construct a simplified geometry of the interface with effective width $\ell$ [Fig.~\ref{fig:interface}(b)]. We recall that the flux $s$ vanishes in bulk, and its integral through the interface also vanishes. Hence, the flux has to cross the nullcline at $\phi_\ast$ within the interface. While the exact value depends on the (yet unknown) details of the interface, within this simplified geometry $\phi_\ast\simeq(\phi^++\phi^-)/2$ is approximated by the arithmetic mean of the coexisting concentrations and $\al_\ast=\bar\al(\phi_\ast)$. We define the rate of events per area
\begin{equation}
    \sig \equiv \IInt{x}{x_\ast}{\infty} s(x)
\end{equation}
with $x_\ast$ the interface position, i.e., $\phi_\ast=\phi(x_\ast)$. The integral of $s(x)$ from $-\infty$ to $x_\ast$ then has to equal $-\sig$ in order for the total charge to vanish. We now replace the interface by two planar sheets with $-\sig$ at $x_\ast-\ell/2$ and $+\sig$ at $x_\ast+\ell/2$ [Fig.~\ref{fig:interface}(c)]. Since the total charge of the interface is zero, outside of the two planes all currents vanish. Inside the interface resembles a capacitor with constant current $\bm j_2=-m\nabla\epsilon=-\sig\bm e_x$. We integrate this relation to obtain
\begin{equation}
    \epsilon^+ - \epsilon^- = \IInt{x}{x_\ast-\ell/2}{x_\ast+\ell/2} \frac{\sig}{m(x)} \simeq \frac{\sig\ell}{m_\ast} \equiv \Delta_0.
    \label{eq:eps:flat}
\end{equation}
The final result follows in the limit of a sharp interface with $m_\ast=m(\phi_\ast)$, where $\ell$ is much smaller than the length on which we resolve the interface [cf. Fig.~\ref{fig:interface}(c)]. The superscripts indicate the bulk phases for $x\ll x_\ast$ ($-$) and $x\gg x_\ast$ ($+$) with order parameters $\phi^\pm$ and $\epsilon^\pm=\bar\epsilon(\phi^\pm)$. In agreement with the electrostatics of a dipole sheet~\cite{jackson21}, the potential difference $\epsilon$ thus jumps across the interface by $\Delta_0$ and the current $\sig$ is proportional to this gap. The rebalancing potential obeys $\nabla\cdot(\tfrac{m}{\al}\nabla\eta)=s$ following from $\nabla\eta+\al\nabla\epsilon=0$ together with Eq.~\eqref{eq:eps}. Performing the same calculation leads again to a jump
\begin{equation}
    \eta^+ - \eta^- \simeq -\al_\ast\frac{\sig\ell}{m_\ast} = -\al_\ast\Delta_0
    \label{eq:eta:flat}
\end{equation}
across the interface eliminating the unknown quantity $\sig\ell$. Rewriting Eq.~\eqref{eq:eta:flat} together with Eq.~\eqref{eq:eps:flat} we find our second central result
\begin{equation}
    \mu^+ - (\al^+-\al_\ast)\epsilon^+ = \mu^- - (\al^--\al_\ast)\epsilon^-
    \label{eq:mu:jump}
\end{equation}
for the chemical potential. We thus seek concentrations $\phi^\pm$ on the nullcline such that the bulk pressure is equal and the bulk chemical potential obeys Eq.~\eqref{eq:mu:jump}.

The picture emerging from these considerations is as follows. In bulk, biasing transitions towards, say, species 2 (i.e., $\Delta\mu>0$) shifts their chemical potential $\mu_2>\mu_1$ up to balance active and passive biases in order to remain on the nullcline with vanishing net flux $s=0$ but non-vanishing dissipation [Fig.~\ref{fig:system}(b)]. If this shift depends on the order parameter $\phi$ then it is different in the coexisting phases. Specifically, Eq.~\eqref{eq:eta:flat} tells us that the chemical potentials of the two species behave as
\begin{equation}
    \mu_1^+ = \mu_1^- - \al_\ast\Delta_0, \qquad \mu_2^+ = \mu_2^- + (1-\al_\ast)\Delta_0.
\end{equation}
At the interface, the system has to compensate the difference through a non-vanishing flux driving particle currents across the interface.

\subsection{Illustration}

To illustrate the general formalism, we turn to a specific example for which we need to specify the free energy density $f(\al,\phi)$ as well as expressions for the transition rates $k^\nu_\pm$. To obtain a specific form for the free energy density, we assume that molecules in conformation 1 are ``inert'', i.e., they do not interact with other molecules. In contrast, molecules in conformation 2 can bind to other molecules in the same conformation releasing the (free) energy $-\eps_\text{b}$ [cf. Fig.~\ref{fig:system}(a)]. Confining molecules to a lattice accounting for the excluded volume, the equilibrium behavior falls into the Ising model class and the phase diagram has been studied, e.g., in the context of molecules adsorbed onto a surface~\cite{picard23}, from which we take the mean-field free-energy density
\begin{equation}
    f(\al,\phi) = -\frac{1}{2}(\al\phi)^2 + \al\phi\Delta f + T\phi f_0(\al) + Tf_0(\phi) + \phi f_1
    \label{eq:free_energy_density}
\end{equation}
with (negative) entropy of mixing $f_0(x)=x\ln x+(1-x)\ln(1-x)$. Here and in the following, free energies and the temperature $T$ are measured in units of $\eps_\text{b}$ (which, for simplicity, we assume to be independent of temperature). Moreover, we rescale concentrations $v_T\phi\to\phi$ so that $0\leqslant\phi\leqslant 1$ is dimensionless. Due to different intramolecular degrees of freedom, the free energy $f_k$ of each conformation might differ with $\Delta f\equiv f_2-f_1$. Interfacial fluctuations in essentially the same model, but chemically driven, have been studied recently~\cite{cho25}.

In addition to the free energy density, we need to specify the transition rates $k_\pm^\nu$. For the active channel, we choose Glauber-like rates
\begin{equation}
    k^\text{a}_\pm(\phi) = k^\text{a}(\phi)\frac{2}{1+e^{\mp\Delta\mu/T}}
\end{equation}
obeying the local detailed balance condition. This functional form for the rates is bounded by the attempt rate $k^\text{a}$ and does not simply increase exponentially as the driving $\Delta\mu$ is increased, taking into account that chemical transformations of macromolecules are typically rate limited. Here we choose $k^\text{a}(\phi)=k^\text{a}_0(1+\zeta\phi)$ as active attempt rate with coefficient $\zeta$ determining the dependence on the total concentration. The difference $\bar\epsilon$ of chemical potentials on the nullcline [Eq.~\eqref{eq:nc:eps}] depends on the ratio $k_0^\text{p}/k^\text{a}_0$ of bare attempt rates with $k^\text{p}_\pm = k^\text{p}_0$, and in the limit $k^\text{p}_0\gg k^\text{a}_0$ of slow active transformations $\bar\epsilon\to0$ whereas in the opposite limit $k^\text{p}_0\ll k^\text{a}_0$ we find $\bar\epsilon\to\Delta\mu$. The implicit condition for the nullcline $\bar\al(\phi)$ reads
\begin{equation}
    \frac{1}{\phi}\left.\pdv{f}{\al}\right|_{\bar\al} = -\bar\al\phi + \Delta f + T \ln \frac{\bar\al}{1-\bar\al} = \bar\epsilon(\phi).
\end{equation}
For constant $\bar\epsilon$, we simply obtain a shifted bias $\Delta f-\bar\epsilon$ that determines the nullcline.

\begin{figure}[h!]
    \centering
    \includegraphics{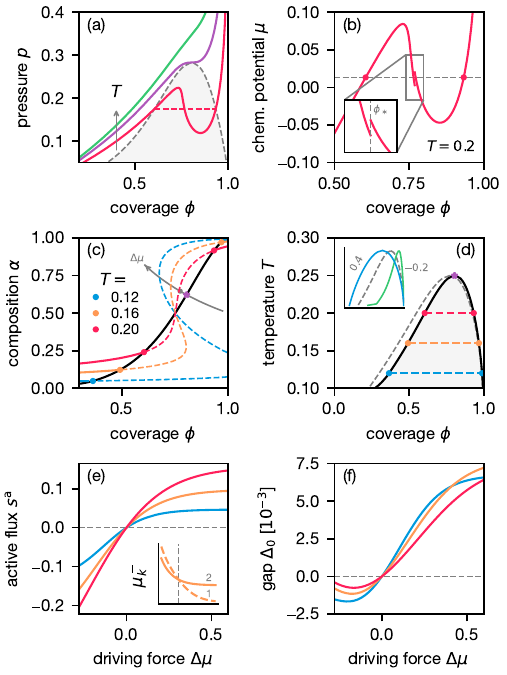}
    \caption{Coexistence and phase diagram in a chemically active binary mixture with non-constant rates ($\zeta=0.1$, $\Delta f=0.35$, $k_0^\text{p}/k^\text{a}_0=1$). (a)~Pressure $p(\phi)$ for three temperatures above, at, and below the critical temperature for $\Delta\mu=-0.05$. The dashed line indicates $p_\text{coex}$ for $T=0.2$. (b)~Chemical potential at $T=0.2$ with the coexisting concentrations $\phi^\pm$. The insight shows the jump at $\phi_\ast$ between the chemical potentials in the dilute and dense regions. (c)~Binodal curve together with nullclines for three selected temperatures. Dashed lines indicate the portion of the nullcline where the uniform system is unstable. Also shown is the position of the critical point and how it changes as the driving force $\Delta\mu$ is increased (solid gray line). (d)~Binodal curve enclosing the two-phase region (shaded) between dilute and dense phase. Quenching a uniform system inside the two-phase region will lead to coexistence with the coexisting concentrations $\phi^\pm$ on the binodal (horizontal dashed tie lines). The gray dashed line shows the equilibrium reference phase diagram for $\Delta\mu=0$. The inset shows two binodals for stronger driving: $\Delta\mu=0.4$ (blue line) and $\Delta\mu=-0.2$ (green line). (e)~Reactive flux through the active channel as a function of driving force $\Delta\mu$ in the dilute phase ($\phi^-$). The inset shows the chemical potentials $\mu^-_k$, which are equal in thermal equilibrium ($\Delta\mu=0$). (f)~The gap $\Delta_0$ as a function of driving force.}
    \label{fig:maxwell}
\end{figure}

We are now in the position to study the impact of non-constant transition rates ($\zeta\neq0$) on phase coexistence. Below a critical temperature $T_\text{c}$, the pressure becomes a non-monotonic function [Fig.~\ref{fig:maxwell}(a)], allowing for the coexistence of two phases with different $\phi^\pm$. Following Eq.~\eqref{eq:mu:jump}, we need to shift the chemical potential in the dilute and dense phase, which leads to a jump at $\phi_\ast$ [Fig.~\ref{fig:maxwell}(b)]. We numerically solve both conditions simultaneously, yielding the coexisting concentrations $\phi^\pm$ with corresponding mole fractions $\al^\pm=\bar\al(\phi^\pm)$. We plot the resulting binodal curve as a function of temperature in chemical space [Fig.~\ref{fig:maxwell}(c)] and in the plane spanned by $\phi$ and $T$ [Fig.~\ref{fig:maxwell}(d)], where the binodal encloses the two-phase region. The shape of the binodal depends on the driving strength $\Delta\mu$. Here we have set $\zeta=0.1$ and for positive $\Delta\mu>0$ coexistence is enhanced: the two-phase region becomes broader and the critical point moves to smaller $\phi$. The dissipation in this regime saturates to a plateau for large $\Delta\mu$ [Fig.~\ref{fig:maxwell}(e)], demonstrating that the mixture is driven beyond the linear response regime. We confirm that biasing towards the interacting species M$_2$ increases its chemical potential $\mu_2>\mu_1$ compared to the inert species [inset of Fig.~\ref{fig:maxwell}(e)]. Consequently, for $\Delta\mu<0$ we find that chemical driving disrupts interactions and thus suppresses coexistence. The binodal now becomes narrower and is pushed to the right [inset of Fig.~\ref{fig:maxwell}(d)] until for some critical driving strength no coexistence is possible anymore and the system becomes a driven homogeneous fluid. The gap $\Delta_0=\epsilon^+-\epsilon^-$ between dilute and dense phase is a non-monotonic function of $\Delta\mu$ [Fig.~\ref{fig:maxwell}(f)]. Since the magnitude and sign of $\Delta_0$ are also determined by $\zeta$, setting $\zeta$ to a negative value inverts these trends with respect to $\Delta\mu$.


\section{Conclusions}

We have studied coexistence in a chemically driven binary mixture and determined the non-equilibrium coexistence criteria. Our analytical results confirm and underpin recent numerical results~\cite{berthin25,cho25,lavagna25}. In contrast to many works also starting from the continuity equations~\eqref{eq:rho} with chemical fluxes, we have leveraged recent advances from stochastic thermodynamics~\cite{falasco25} to strictly enforce local detailed balance also on the level of fluxes [Eq.~\eqref{eq:ldb}]. This condition ensures the thermodynamic constraint that the dissipation rate equals the molecular solute flux between reservoirs times the free energy liberated per event [Eq.~\eqref{eq:diss}]. As a consequence of this constraint, the difference $\epsilon$ between the chemical potentials of macromolecular species on the nullcline (a thermodynamic quantity) is entirely determined by the kinetics [Eq.~\eqref{eq:nc:eps}]. If transition rates do not depend on space then the conventional Gibbs criteria of equal pressure and rebalancing (instead of chemical) potential are recovered to determine coexisting concentrations even though the system is globally driven into a non-equilibrium steady state. The reason is that fluxes in chemical space still balance locally and all particle currents vanish.

On the other hand, concentration-dependent rates force the difference $\bar\epsilon(\phi)$ between chemical potentials to be different in the dense and dilute phase, which in turn forces the net flux to be non-zero [Eq.~\eqref{eq:eps}] and thus non-vanishing particle currents emerge. These currents are confined to the interface while the total current still has to vanish in steady state. Here we have shown that generalized coexistence criteria hold, in which bulk chemical potentials are no longer equal in the coexisting phases but shifted in order to balance the currents across their interface. The solution is formally equivalent to the jump of the electric potential across a dipole sheet in electrostatics. This jump depends on features of the interface, which in a first step we have treated in a simplified sharp-interface approximation.

Intriguingly, a jump of the (effective) chemical potential has also been found to determine phase coexistence in a scalar active field theory~\cite{tjhung18,cates25,hertag26}. These theories augment a particle current driven through a free-energy gradient by non-potential terms, which can be absorbed into an \emph{effective} free energy through an integrating factor~\cite{wittkowski14,solon18,solon18a,speck21a,omar23}. Indeed, the large-scale dynamics of ideal reaction-diffusion systems with a conservation law can be cast into the form of a scalar active field theory~\cite{robinson25}. It will be illuminating to further explore these formal analogies between chemically driven mixtures and active matter.


\begin{acknowledgments}
    We acknowledge financial support by the Deutsche Forschungsgemeinschaft (DFG) through the collaborative research centers TRR 146 (grant no. 233630050) and SFB 1551 (grant no. 464588647). EM thanks the International Max-Planck Research School for Intelligent Systems (IMPRS-IS) for support.
\end{acknowledgments}


%

\end{document}